\documentstyle[aps,epsf,twocolumn]{revtex}
\newcommand{\mfrac}[2]{\mbox{$\frac{#1}{#2}$}}
\newcommand{\p}{{\bf p}}
\newcommand{\bd}[1]{{\bf #1}}

\newtheorem{theorem}{Theorem}

\newtheorem{lemma}[theorem]{Lemma}

\begin{document}
\title{Scalar Field Cosmological Models With Hard Potential Walls}
       
\author{{\bf Scott Foster}}
\date{
Department of Physics and Mathematical Physics\\
The University of Adelaide, Adelaide, Australia 5005.}

\maketitle

\begin{abstract}
The global behavior of scalar field cosmological models with very hard potential walls is investigated via the simple example of an exponentially steep potential well. It is found that the solutions exhibit a non-trivial oscillatory behavior in the approach to an initial space-time
singularity. This behavior can be interpreted as being due to the inability of the scalar field to negotiate the walls of the steep potential well.     
\end{abstract}

\section{Introduction}
According to the inflationary universe scenario \cite{guth,linde,blau} the matter content in the very early universe can be  modeled by a scalar field $\phi$
with a non-negative self interaction $V(\phi)$. Inflation may be understood physically in terms of the stress energy tensor of the scalar field which is equivalent to a perfect fluid with energy density $\rho$ and pressure $p$ given by 
\begin{equation}\label{rhop}
\rho = -\frac{1}{2}\partial^\mu\phi\partial_\mu\phi + V(\phi)\hspace{1.5cm}
p =-\frac{1}{2}\partial^\mu\phi\partial_\mu\phi -  V(\phi).
\end{equation} 
When $V$ becomes large compared to the gradient of the scalar field, the pressure becomes large and negative and the fluid satisfies the approximate equation of state $\rho = -p$ leading to a rapid exponential expansion of space-time. 
This rapid expansion is associated with the fact that gravity becomes
 repulsive in the presence of large negative pressures. Such  equations of state  have traditionally been forbidden in general relativistic matter fields  by the strong energy condition  which, for a perfect fluid, is equivalent to the following familiar constraint on the pressure and energy \cite{hawkandell}:
\[
\rho + 3p > 0.
\]
Thus, for a scalar field to obey the strong energy condition we would require that
\[
-\partial_\mu\phi\partial^\mu\phi -  V(\phi) > 0.
\]
 The fact that scalar fields naturally violate the above inequality means that the singularity theorems \cite{hawkandell} can not be strictly applied when such fields are present in the early universe. Indeed, it is relatively simple to construct exact singularity free scalar field cosmologies (eg \cite{mad2}). Other exact scalar field solutions are known which possess singularities, but have no particle horizons \cite{lucmat}, indicating that scalar field cosmologies can exhibit a diverse range of behavior.

It turns out however, that all of the known singularity free and horizon free solutions are unstable and it can be demonstrated  that, provided the potential  diverges slower than exp$(\sqrt{6}\phi )$ as $\phi\rightarrow \pm\infty$,  almost all spatially flat Friedmann Robertson Walker (FRW) cosmologies have initial singularities. Furthermore, they have asymptotic equation of state $\rho =p$ and may be asymptotically approximated by the general solution for the massless scalar field \cite{flatpap}.    
      
 However, since we have no direct emperical data with which to determine the detailed nature of  physical fields
 at energy densities comparable with those of the early universe, we should not rule out the possibility that  scalar fields present in the early universe might have much  steeper potentials.      
   
It is therefore natural to ask whether singularity free or horizon free
 cosmologies can arise as a result of very steep potentials. In particular,
 we might expect
that  a very steep potential well could inhibit the divergence of the scalar 
field, thereby slowing down the gravitational collapse and
 resulting in singularity 
or particle horizon avoidance.

 Indeed, it can easily be demonstrated that  for very steep potentials $V$  can not be neglected asymptotically in the past as can be done with less steep potentials. This can be seen as follows: 

For simplicity, we confine attention to the FRW line element 
\begin{equation}\label{eq:rw}
 ds^2 =- dt^2 + a(t)^2\sum_{i=1}^3 (dx^i)^2.
\end{equation}   
We define the expansion $K(t)$ as the function
\begin{equation}\label{eq:adot}
K=3\frac{\dot{a}}{a},
\end{equation}
where the ``dot'' indicates differentiation with respect to time $t$. K may be interpreted physically as the rate of expansion of the spatial volume element $v=a^3$.

The evolution equations governing the interaction of a general scalar field with the above metric are \cite{blau}:
\begin{eqnarray}
\ddot{\phi}& =  & -K \dot{\phi}- V'(\phi)\label{eq:S1}\\
\dot{K}& = & -\mfrac{3}{2}\dot{\phi}^2 \nonumber
\end{eqnarray}
\begin{equation}\label{algcon}
K^2 =  3V(\phi) + \frac{3}{2}\dot{\phi}^2.
\end{equation}
The general solution for the massless scalar field is obtained by setting $V=0$ in the above and solving to obtain. 
\begin{equation}\label{eq:massless}
 K = \frac{1}{t}\hspace{1.5cm} \phi = \pm\sqrt{\mfrac{2}{3}}\ln\frac{t}{c}
\end{equation}
Let us now use (\ref{eq:massless}) as an approximation to the solution for arbitrary $V$ in the limit $t\rightarrow 0$. Substituting (\ref{eq:massless})
 into (\ref{algcon}) we obtain 
\[
\frac{1}{t^2} = \frac{1}{t^2} + V\left(\pm\sqrt{\mfrac{2}{3}}\ln\frac{t}{c}
\right) + h
\]   
where,  $h$ indicates terms of higher order in $t$. For  this 
expression to be consistent the following must be true:  
$$
\lim_{t\rightarrow 0} t^2V\left(\pm\sqrt{\mfrac{2}{3}}\ln\frac{t}{c}\right)=0.
$$
i.e.
$$
\lim_{\phi\rightarrow \pm\infty}e^{-\sqrt{6}|\phi|}V(\phi )=0.
$$ 
In other words, in order for the massless scalar field to represent a consistent 
asymptotic approximation of a scalar field model it is  necessary that the potential diverge slower than
${\displaystyle e^{-\sqrt{6}|\phi|}}$. Exponential potentials therefore 
seem to accommodate  a kind of transition between soft wall potentials
 and some 
other  qualitative  regime. 

In this paper  we examine
the behavior of very steep potentials by means of a simple and interesting example.
 
\section{The Model and Dynamical Equations.}
 The simplest example of a potential which is exponentially steep at both infinity and negative infinity is an exponential
well of the form.
\[
  V(\phi)=ae^{\lambda\phi} + be^{-\mu\phi}. \]
where $a, b, \lambda$ and $\mu$ are positive constants.
 In the subsequent analysis we will confine attention to the case where $a=b=1$ and $\lambda=\mu$ so that
\begin{equation}
V(\phi)=e^{\lambda\phi} + e^{-
\lambda\phi}.  \label{eq:epot}
\end{equation}
By increasing $\lambda$ we shall be able to 
 investigate how the qualitative behavior
 of the system changes as the potential well becomes increasingly steep.
The analysis for the more general potential is similar and the conclusions
 are essentially the same. 

We are interested only in expanding solutions of (\ref{eq:S1}) so we may   replace $K$ and $\dot{\phi}$ with the new set of coordinates
\begin{equation}\label{var1.5}
x=\frac{1}{K}\hspace{1cm}y=\sqrt{\mfrac{3}{2}}\frac{\dot{\phi}}{K}
\end{equation}
and introduce a new time coordinate
 \begin{equation}
\label{taudef}
\tau=\ln v(t) + \tau_0
\end{equation}
where $v=a^3$ is the spatial volume element and $\tau_0$ is some constant.
 $\tau$ is well defined since $v$ is strictly increasing. Furthermore, $v$ 
goes to zero if and only if $K$ goes to infinity \cite{hawkandell} so (\ref{taudef}) implies that $K\rightarrow \infty$ as $\tau\rightarrow 
-\infty$. Differentiating (\ref{taudef}) we find (recalling that $K=\dot{v}/v$)
 
\begin{equation}
\frac{d}{dt}=K\frac{d}{d\tau}.\label{eq:tau}
\end{equation}

In terms of these coordinates the field
equations (\ref{eq:S1}) with the potential (\ref{eq:epot}) may be written as a dynamical system:  
\begin{eqnarray}
\frac{dx}{d\tau}&=&y^2x \nonumber\\
\frac{dy}{d\tau}&=&-y- 3\alpha x^2(e^{\sqrt{6}\alpha {\phi}}-e^{-\sqrt{6}\alpha {\phi}})+ y^3 \label{eq:S1exp}\\
\frac{d{\phi}}{d\tau}&=&y \nonumber
\end{eqnarray}
where $\alpha=\frac{\lambda}{\sqrt{6}}$. The ``constraint'' equation (\ref{algcon}) becomes
\begin{equation}\label{eq:conexp}
y^2+3x^2(e^{\sqrt{6}\alpha {\phi}}+e^{-\sqrt{6}\alpha {\phi}})=1.
\end{equation}
We can simplify the equations by  defining  the variables
\begin{equation}\label{eq:pqdef}
p=\sqrt{3}e^{-\sqrt{\frac{3}{2}}\alpha {\phi}}x\hspace{1.5cm} q=y.
\end{equation}
then the constraint equation may be written
\begin{equation}\label{eq:conexp2}
p^2+q^2=1-p^2e^{2\sqrt{6}\alpha {\phi}}.
\end{equation}
Substituting (\ref{eq:pqdef}) and (\ref{eq:conexp2}) into (\ref{eq:S1exp}) 
we obtain 
\begin{eqnarray}
\frac{dp}{d\tau}&=&-\alpha pq+ pq^2\nonumber\\
\frac{dq}{d\tau}&=&q^3+\alpha q^2 -q-\alpha + 2\alpha p^2.\label{eq:S2exp}
\end{eqnarray}
These equations constitute a 2-dimensional dynamical system on the $p$-$q$ plane.
 We may define the physical
phase space $\Omega$, according to (\ref{eq:conexp2}),
 as the set $\Omega=\{(p,q):p^2+q^2 < 1, p>0\}$. In other words,
 all physical trajectories lie on the interior of the unit disc to the
right of the $q$ axis. The unphysical boundary $\partial\Omega$ is a closed
 curve consisting of the union of the smooth arc
$\partial\Omega_1 = \{(p,q): p^2+q^2=1, p>0\}$ and the
line segment $\partial\Omega_2 =\{(p,q): p=0, |q|\leq1\}$. The
asymmetric appearance of $\partial\Omega$ is a consequence of the
positive exponential term in the definition of the coordinate $p$ and is not
a physical property of the system itself. In fact it should be pointed out
that $\partial\Omega_1$ maps onto $\partial\Omega_2$ under the transformation
$(\phi ,\dot{\phi})\mapsto -(\phi ,\dot{\phi})$.
$\partial\Omega$ corresponds to the infinity of the expansion $K$.
 In order to see
 this define the function
\begin{equation}\label{eq:Hdef+}
H(p,q)=p^2(1-p^2-q^2)
\end{equation}
Clearly $H$ is strictly positive everywhere on $\Omega$ and vanishes
 identically on $\partial\Omega$. From
(\ref{eq:conexp2}) and (\ref{eq:pqdef}) we see that 
\begin{equation}  \label{eq:hx}
H=p^4e^{2\sqrt{6}\alpha {\phi}}=9 x^4. 
\end{equation}
Thus $\partial\Omega$ is just the set of all points for which $x=0$ which is
 by definition
the infinity of $K$. Evaluating the directional
derivative of $H$ along the flow  using (\ref{eq:S2exp}) we find
\begin{equation}\label{eq:hdiff}
\frac{dH}{d\tau}=4H q^2
\end{equation}
which is non-negative everywhere on the interior of $\Omega$ .
In fact the derivative of $H$ with respect to $\tau$ is strictly positive
 everywhere on the interior of $\Omega$ except where $q=0$. Observe
also that ${\displaystyle \frac{dH}{d\tau}=0}$ on $\partial\Omega$ indicating that 
the boundary is an invariant manifold (tangent to the flow). No physical
trajectories can therefore cross $\partial\Omega$ into the unphysical 
domain beyond.  Since $\Omega$ is compact all trajectories must
possess an  past-limit set which is invariant under the flow.
 Since $H$ is monotonic, any limit
point must have ${\displaystyle \frac{dH}{d\tau}=0}$.
Therefore, all limit sets must be subsets of either $\partial\Omega$ 
or the line $q=0$. From  (\ref{eq:S2exp}) the only invariant
subset of $q=0$ is the equilibrium 
 point $\p_d=
(\frac{1}{\sqrt{2}}, 0)\footnote[2]{In order to avoid confusion points on the 
$(p,q)$ plane will be labeled in bold  print.}$.
 However this point is a local maximum of
$H$ as is easily verified by evaluating its gradient. Since 
$H$ is monotonic increasing, no solutions can be past asymptotic to
$\p_d$ other than the steady state solution on $\p_d$ itself.
It follows that the past-limit sets of all other solutions lie
on the boundary $\partial\Omega$.

By the time reverse of the above argument it is clear that all solutions,
with the exception of those unphysical solutions lying on $\partial\Omega$
are future asymptotic to $\p_d$.

The future asymptotic set $\p_d$ represents a vacuum de Sitter space-time 
with constant expansion $K=\sqrt{6}$ and $\phi$ identically zero. This is
consistent with what we would expect for a scalar field cosmology with 
non-zero vacuum energy.
 
  To summarize, it has been established that (almost) all solutions
originate near $\partial\Omega$ and subsequently evolve
towards the global attractor $\p_d$ (which represents de Sitter space) as
$t\rightarrow \infty$. Let us now examine the behavior of the system on
 and near
$\partial\Omega$ in more detail.

\section{The Behavior Close to $\partial\Omega$. }

 There are a maximum 4 equilibrium points of 
(\ref{eq:S2exp}) lying on $\partial\Omega$. These are $(\sqrt{(1-\alpha^2)},
\alpha )$, $(0,1)$, $(0,-1)$ and $(0,-\alpha)$, which we label
$\p_1$, $\p_+$, $\p_-$ and $\p_2$ respectively. Observe that $\p_1$ and $\p_2$
 only exist as distinct equilibrium points on
$\partial\Omega$ when $\alpha<1$. For values of $\alpha$ greater than or
 equal to  one, $\p_\pm$ are the only equilibrium points.
In order to investigate the behavior close to $\p_{\pm}$ equations
(15) and (16) can be linearized about these points. The linearized system is:
\begin{eqnarray}\label{eq:cornlin}
\frac{dp}{d\tau}&=&(1\mp\alpha) p\\
\frac{dq}{d\tau}&=& 2(1\pm\alpha)(q\mp 1)
\end{eqnarray}
The solution to the linear system is
\begin{equation}\label{eq:cornapp}
p=p_0e^{(1\mp\alpha)\tau} \hspace{1.5cm} q=\pm 1\mp \delta_0e^{2(1\pm\alpha)\tau}
\end{equation}
where $p_0$ and $\delta_0$ are positive constants.
\subsection{The Flow For $\alpha <1$.} 
\begin{figure}[htb]
\epsfysize=7cm
\epsfxsize= 7cm
\begin{center}
\leavevmode
\epsfbox{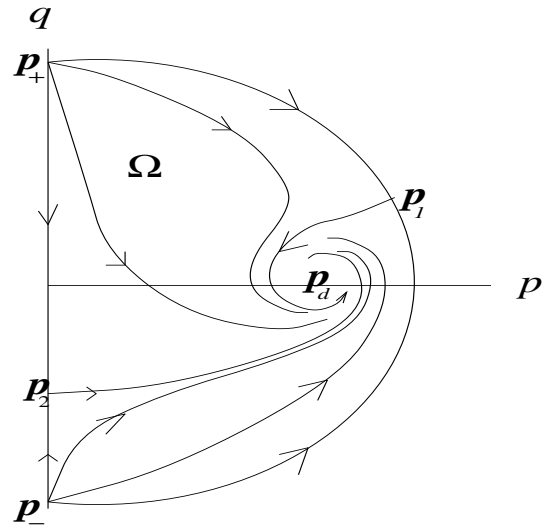}
\end{center}
\caption{Sketch of $\Omega$ 
 showing all possible  equilibrium points, direction
of the flow on the boundary, and some typical trajectories. Trajectories
on the boundary are future asymptotic to $\p_1$ and $\p_2$ but all 
trajectories on the interior approach $\p_d$.}
\label{fig:exp}\end{figure}

 When $\alpha<1$ both
exponential terms have positive coefficients indicating  that
 $\p_{\pm}$ are sources of the linear system and therefore, by the 
Hartman-Grobman Theorem \cite{gucholms,wigg}, of the non-linear system also.
Since the unphysical solutions on $\partial\Omega$ move away 
from $\p_{\pm}$ they must asymptotically approach $\p_1$ and $\p_2$ (in the forwards time sense). However all
{\em physical} solutions (ie, those on the interior of $\Omega$) asymptotically approach $\p_d$ in the future. $\p_1$
and $\p_2$ must therefore be unstable  saddles.
 This can be seen more clearly 
from inspection of the geometry of the flow as illustrated in Fig. 
\ref{fig:exp}.\\ ( An alternative way to verify that $\p_1$ and $\p_2$
are saddles  is  by linearizing 
(\ref{eq:S2exp}) about these points but this exercise shall be left to
 the reader who remains unconvinced).
 With the exception of the 2 solutions
past asymptotic to $\p_1$ and $\p_2$ respectively, and the steady state
solution lying on $\p_d $, all physical trajectories must asymptotically
approach either $\p_+$ or $\p_-$ as $\tau\rightarrow -\infty $, as indicated
 in Fig. \ref{fig:exp}.

Using the fact that $1-q^2$ decays exponentially to zero as $\tau\rightarrow 
-\infty$ we may integrate (\ref{eq:hdiff}) to obtain a first order 
expression for $H$:
\[ H=H_0e^{4\tau}
\]
and hence
\[
x=x_0e^\tau. \]
Using (\ref{eq:tau}) we obtain a first order expression for $t$: 
\[t=x_0e^\tau. \]
Observe that $t\rightarrow 0$ as $\tau\rightarrow -\infty$ indicating that
$\p_\pm$  correspond to  space-time singularities. 
Using the definitions of $x$ and $q$ we thus obtain the asymptotic solution
for $K$ and $\dot{\phi}$ in the neighbourhood of $t=0$:
\begin{equation}
K=\frac{1}{t}\hspace{1.5cm} \dot{\phi}=\pm\sqrt{\frac{2}{3}}\frac{1}{t}
\end{equation}
and upon integration of $\dot{\phi}$
\begin{equation}
\phi=\pm\sqrt{\frac{2}{3}}\ln \frac{t}{c}.
\end{equation}
This is, of course, the general solution for the massless
scalar field. We thus conclude that when $\alpha<1$, ie $\lambda<\sqrt{6}$,
the potential $V$ is not dynamically significant near the singularity.
This is precisely the behavior we would have expected
 from the simple 
calculation in Section 1 for potentials that go to infinity slower
than $\exp (\sqrt{6}\phi)$. In fact the topological structure of solution space is typical of inflationary models.  All s
solutions
emerge from the sources $\p_\pm$ except for the 
 seperatrices $I_1$ or $I_2$ which we define as 
the unique solutions past 
asymptotic to the saddles $\p_1$ and $\p_2$ respectively. 
We may infer that these
solutions characterize the inflation in the system. (It can be 
easily verified that the behavior
close to $\p_1$ and $\p_2$ is approximated by the 
 exact power law cosmologies \cite{lucmat}. However this 
behavior  only corrosponds to inflation
when $\alpha^2<\frac{1}{3}$ since this condition 
ensures that the square of the
$q$ coordinate at $\p_+$ and $\p_-$ is less than $\frac{1}{3}$ which is 
necessary and sufficient for violation of the strong energy condition.)

\subsection{The Flow For $\alpha \geq 1$}
Let us now consider what happens when $\alpha \geq 1$. As
 we demonstrated in Section 1, the potential in this
 case becomes too steep for the solution with $V=0$ to be consistent as an
asymptotic solution.
 In more physical terms, the gravitational
 expansion is unable to dominate the scalar field self interaction in the 
initial expansion phase of the universe. For expediency, we shall assume in 
what follows that $\alpha>1$. The special case $\alpha=1$ differs
 in the details
 of analysis, but the qualitative features of the solutions turn out to be
essentially the same. 

The only  equilibrium points possessed by the system when $\alpha>1$
 are $\p_+$, $\p_-$ and
the sink $\p_\infty $
. Inspection of (\ref{eq:cornapp}) reveals that $\p_\pm$ become hyperbolic
 saddles in 
this regime.

\begin{figure}[htb]
\epsfysize=7cm
\epsfxsize= 7cm
\begin{center}
\leavevmode
\epsfbox{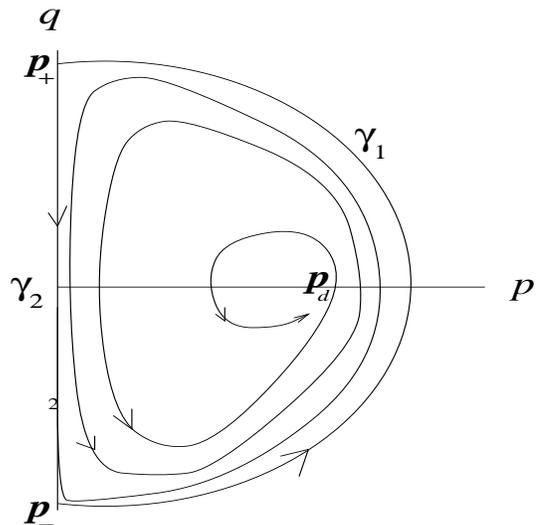}
\end{center}
\caption{Sketch of $\Omega$ for $\alpha > 1$ 
showing the solutions on the  boundary and a typical solution on the interior.
The points $\p_\pm$ are saddles and therefore unstable, but solutions 
must approach the boundary so the union of solutions on $\partial\Omega$
forms a limit cycle.}
\label{fig:exp2}\end{figure}

The only solution past asymptotic to $\p_-$ is the unphysical trajectory 
$\gamma_1$ which emerges from $\p_-$ at $\tau=-\infty$ and proceeds anticlockwise along $\partial\Omega_1$ (the unit circle), reaching $\p_+$ at
 $\tau=\infty$. Similarly,
the only solution which originates at $\p_+$ is the unphysical
trajectory $\gamma_2$ which emerges from $\p_+$ at $\tau=-\infty$ and proceeds
along $\partial\Omega_2$ (the $q$-axis), approaching $\p_2$ asymptotically as
 $\tau\rightarrow\infty$.
 The
union of $\gamma_1$, $\gamma_2$ 
 is a closed heteroclinic cycle $\gamma_L$ on
 $\partial\Omega$ which forms a limit cycle for trajectories on the interior
of $\Omega$. That is, as $\eta\rightarrow -\infty$ a typical trajectory will
approach $\partial\Omega$, spiraling clockwise (in the reverse time  sense) an
 infinite number of times. The solution will asymptotically
approach the non-physical solutions
$\gamma_1$ and $\gamma_2$ but unlike these will always avoid the equilibrium
points $\p_+$ and $\p_-$ and will continue to spiral around $\partial\Omega$
 {\it ad infinitum}. Fig \ref{fig:exp2}.

In order to interpret these results recall firstly that,  from
(\ref{eq:Hdef+}) and (\ref{eq:hdiff}), $H$ monotonically decreases to 0
as $\tau\rightarrow -\infty$ and hence the expansion $K$ diverges 
monotonically to infinity. The asymptotic periodic behavior of the trajectories
must therefore represent oscillations of the scalar field $\phi$.
 Inspection of 
(\ref{eq:conexp2}) indicates that $\phi=-\infty$ on the semi-circle
 $\partial\Omega_1$.
It follows from symmetry that $\phi=\infty$ on $\partial\Omega_2$. As the
gravitational expansion diverges to infinity ( and correspondingly
the scale parameter $a$ tends to 0) the scalar field oscillates about
$\phi=0$ with the amplitude of oscillation increasing for each successive
cycle, asymptotically diverging to infinity.

No asymptotic equation of
state is satisfied by the scalar field.
In the ``corners'' of $\Omega$, close to the points $p_{\pm}$,
 the equation of state
is approximately $\rho=p$ since $\phi^2\simeq \frac{2}{3}K^2$ which implies,
via (\ref{algcon}) that $\dot{\phi}^2>>V(\phi)$. 
 This corresponds to the period where the scalar field is close to 
the bottom of the potential well. 

Close to the $q$-axis, on the other hand,
the equation of state is approximately $\rho=-p$ since $\dot{\phi}\simeq 0$
whilst $V(\phi)\geq 2$. This corresponds to
the scalar field reaching the top of its roll before  
accelerating down the potential well again. The asymptotic behavior 
of the space-time may therefore be characterized by an infinite sequence of
``stiff'' phases, punctuated by de Sitter phases. During the 
de Sitter phase the matter violates the strong energy condition.  

\section{The Existence of a Singularity and Particle Horizon}
  
 Since an infinite number of physical cycles of the scalar field occur
for the case $\alpha >1$, 
it might 
reasonably be  expected that an infinite interval of proper time must also
 elapse since the oscillations of the 
scalar field could be used as a natural physical clock which would
measure
an infinite time interval to the past of any space-time point. 
If this were the case then the space-time would be non-singular
for all physical solutions.  
It turns out however, that 
the oscillations pile up on each other and a space-time singularity is 
indeed reached after a finite time interval.

In order to show this we must estimate the proper time that elapses to
the past of some arbitrary initial point $\p_0$, on the trajectory
$\psi_{\p_0}(\tau)$. Since all trajectories are asymptotically tangent
to the boundary $\partial\Omega$ it will be sufficient to restrict attention
to initial points lying in the set
 $\Sigma_{0}=\{(p,q): q=1-\epsilon,\; 0<p\leq\epsilon\}$ where
 $\epsilon >0$ may be chosen arbitrarily small.

Before proving that space-time singularities and particle horizons exist for
all solutions it will be convenient to prove the following lemma;
\begin{lemma}\label{lem:1}
For all $ 0<n<1$ there exist positive numbers
 $\epsilon$ and $p_m$ such that if $x_0=(p_0, 1-\epsilon)$ is in 
$\Sigma_{0}$ and $p_0<p_m$ then 
\begin{equation}\label{hntau}
H(\psi_{\p_0}(\tau))<H_0e^{4n\tau}
\end{equation}
to the past of $\p_0$, where $\psi_{x_0}(\tau)$ is the unique trajectory
 of (\ref{eq:S2exp}) passing through $\p_0$ with $\psi_{\p_0}(0)=\p_0$.
\end{lemma}
Proof:
 Integrating
(\ref{eq:hdiff}) backwards in time from $\tau =0$ to some earlier time $\tau_f$
 we obtain
\begin{equation}\label{eq:htime}
H(\tau_f)=H_0\exp\left[-4\int_{\tau_f}^0 q^2 d\tau \right].
\end{equation}
Let $\Sigma_{ 1}=\{(p,q)\in\Omega:p=\epsilon , \; 1-q\leq\epsilon\}$.
\begin{figure}[htb]
\epsfysize=7cm
\epsfxsize= 7cm
\begin{center}
\leavevmode
\epsfbox{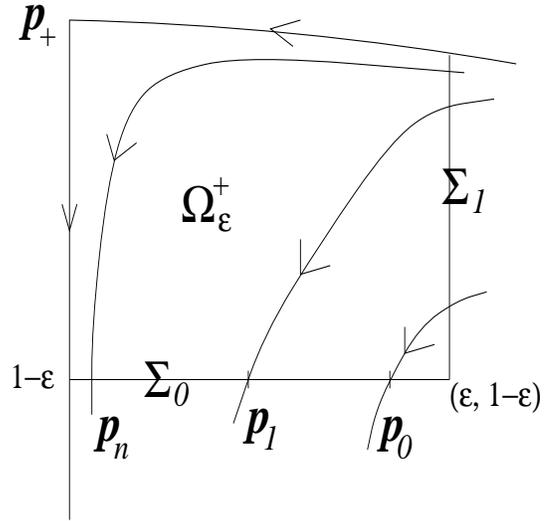}
\end{center}
\caption{The box $\Omega_\epsilon^+$. On each successive cycle 
in the reverse time direction the trajectory
intersects the box for a finite period of time coming progressively closer
to the boundary of $\Omega$ (the arrows indicate the
direction of the flow in the {\em forward} time sense ).}
\label{fig:box}\end{figure}
 As can be seen
from Fig. \ref{fig:box},
 the set $\Sigma_1+\Sigma_0$ encloses a box, $\Omega_\epsilon^+$,
 of area $\simeq \epsilon^2$
in $\Omega$ around $\p_+$. Another box, $\Omega_\epsilon^-$, can similarly be
 constructed around $\p_-$ by
defining the sets  $\Sigma_2=\{(p,q)\in\Omega:p=\epsilon \; 1+q\leq\epsilon\}$
and $\Sigma_3=\{(p,q): q=-1+\epsilon,\; 0<p\leq\epsilon\}$.

Let $I$ be the time interval $[\tau_f, 0]$, then for a given
trajectory $\psi_{\p_0}$ we may write
$I= I_c\cup I_b$ where $I_c=\{ \tau\in I : \psi_{\p_0}(\tau)\in\Omega^\pm_\epsilon \}$ and $I_b=\{ \tau\in I : \psi_{\p_0}(\tau)\not\in\Omega^\pm_\epsilon \}$.
We thus have
\begin{eqnarray}
\int_{\tau_f}^0 q^2 d\tau &=&\int_{I_c} q^2 d\tau +\int_{I_b} q^2 d\tau\\
                         &>&\int_{I_c} q^2d\tau.
\end{eqnarray}
From the definition of $I_c$,
\[
\int_{I_c} q^2 d\tau =\int_{I_c} (1 - O(\epsilon))d\tau.
\]
Fix $n$ and let $m$ be any number satisfying $n<m<1$. 
Then for 
  $\epsilon$  sufficiently small we have;
\begin{equation}\label{int:pis}
\int_{\tau_f}^0 q^2 d\tau > m\int_{I_c} d\tau.
\end{equation}
In order to complete the proof we must show that by choosing $p_0$ 
sufficiently small, the ratio of $\int_{I_c}d\tau$  to
$\int_{I}d\tau$ may be made arbitrarily close to 1.

According to (\ref{eq:S2exp}) $p$ evolves according to the equation
\begin{eqnarray}
\frac{d}{d\tau}\ln p& = &-\alpha q+ q^2\nonumber\\
                     & <& 1 + \alpha
\end{eqnarray}
since $|q|<1$. Therefore if $\tau_2<\tau_1$  we have
\begin{equation}\label{pratbd}
\ln\frac{p_1}{p_2} < (1+\alpha )(\tau_1 -\tau_2),
\end{equation}
where $p_1=p(\tau_1), p_2=p(\tau_2)$.  
The parameter time $\Delta\tau$ needed for a point $\p_0=(p_0,1-\epsilon)
\in\Sigma_{0}$ to flow through $\Omega_\epsilon^+$ and
 reach $\Sigma_{1}$ must therefore satisfy
\begin{equation}\label{deltatau}
|\Delta\tau|>\frac{1}{1+\alpha}\ln\left(\frac{\epsilon}{p_0}\right).
\end{equation}
The modulus sign is necessary because we are tracing 
the trajectory backwards in time so $\Delta\tau$ is negative.
 As $p_0\rightarrow
0$ , $\Delta\tau\rightarrow -\infty$. Thus $\Delta\tau$ may be made
 arbitrarily large by choosing $p_0$ sufficiently small.

\begin{figure}[htb]
\epsfysize=7cm
\epsfxsize= 7cm
\begin{center}
\leavevmode
\epsfbox{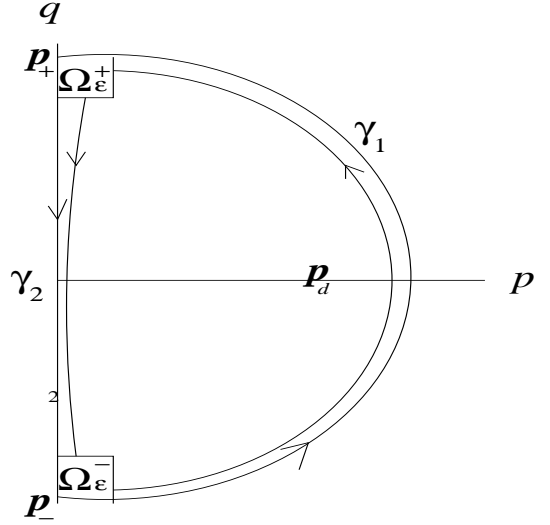}
\end{center}
\caption{The parts of the solution flowing between the boxes
 $\Omega_\epsilon^+$ and $\Omega_\epsilon^-$ approach the restrictions
of $\gamma_1$ and $\gamma_2$ to finite time intervals.}
\label{fig:lim}\end{figure}

 Since $\psi_{\p_0}$
is past asymptotic to the limit cycle $\gamma_l$, the
intersection of the past orbit $O^-_{\p_0}$ with the line $q=1-\epsilon$
(which contains $\Sigma_0$) must
contain an infinite number of points in addition to $\p_0$ itself.
 If $\p_1$ is any such point then
substituting $q=1-\epsilon$ into (\ref{eq:Hdef+}) and using the monotonicity
of $H$  (\ref{eq:hdiff}) we must
 have $p_1<p_0$, where $p_1$ and $p_0$ are the p-coordinate
values of $\p_1$ and $\p_0$ respectively. In other words $\psi_{\p_0}$ must
intersect $\Sigma_{\epsilon 0}$ again after 1 complete cycle of 
$\partial\Omega$ and the point of  intersection $\p_1$ must have a
$p$-coordinate value $p_1$ which is smaller than $p_0$. It follows
that $\psi_{\p_0}$ passes through the box $\Omega_\epsilon^+$ on each
successive cycle and the time interval $\Delta\tau_n^+$ to traverse
$\Omega_\epsilon^+$ on the $n^{th}$ cycle obeys the inequality;
\begin{eqnarray}
|\Delta\tau_n^+|&=& \frac{1}{\alpha +1}\ln\left(\frac{\epsilon}{p_n}\right)\\
                &\geq&\frac{1}{\alpha +1}\ln\left(\frac{\epsilon}{p_0}\right)
\label{in:1}\end{eqnarray}
Similarly,  defining $\Delta\tau_n^-$ to be the time taken for
  $\psi_{\p_0}$ to traverse the box  $\Omega_\epsilon^-$
 on the $n^{th}$ cycle we find from (\ref{pratbd}) using an identical
argument to that above that
\begin{equation}
|\Delta\tau_n^-|\geq \frac{1}{\alpha+1}\ln\left(\frac{\epsilon}{\tilde{p}_n}\right).
\end{equation}
where $\tilde{p}_n$ is the $p$ coordinate of the intersection of $\psi_{\p_0}$
with $\Sigma_3$ on the $n^{th}$ cycle (by the $n^{th}$ cycle I mean, precisely,
one complete circuit from $(p_n-1, 1-\epsilon)\in \Sigma_0$ to $(p_n,
 1-\epsilon )\in\Sigma_0$). Recalling the definition of $H$ 
and using the fact that $q^2$ takes the same value
 on $\Sigma_0$ and $\Sigma_3$
we have 
 for any point on $\Sigma_0$ {\em or} $\Sigma_3$ that $H^2 = 
(2\epsilon - \epsilon^2 )p^2 + O(\epsilon p^4)$. Since $H$ is monotonic
it follows that for $\epsilon$ chosen sufficiently small $\tilde{p}_n<p_0$.
Therefore, 
\begin{equation}\label{in:2}
|\Delta\tau_n^-|\geq \frac{1}{\alpha+1}\ln\left(\frac{\epsilon}{{p}_0}\right).
\end{equation}  
\\
Let us now consider the parameter time $\Delta\tau$
 taken for a point on $\Sigma_1$ to reach
  $\Sigma_2$ under (\ref{eq:S2exp}), close to the semi-circular 
boundary $\partial\Omega_1$. That is , the time taken to flow backwards in
time from
the top box $\Omega_\epsilon^+$ to the bottom box $\Omega_\epsilon^-$
(Fig. \ref{fig:lim}).
 By continuity, $\Delta\tau$ approaches the (negative) 
parameter time interval which the asymptotic solution $\gamma_1$ takes to 
map the point $(\epsilon, \sqrt{1-\epsilon^2})\in\Sigma_1$ to the point
$(\epsilon, -\sqrt{1-\epsilon^2})\in\Sigma_2$. This interval is finite. 
 
Therefore, the time interval to go from $\Sigma_1$ to $\Sigma_2$ must
approach a finite (negative) limit. Its modulus must therefore possess a
finite upper bound $\tau_{1}$. 

Let $\Delta\tau_n^1$ be
the time taken for $\psi_{x_0}$ to go from
 $\Sigma_1$ to $\Sigma_2$ on the $n^{th}$ cycle. Then we have
\begin{equation}\label{in:3}
|\Delta\tau_n^1|<\tau_{1}
\end{equation}
Similarly, if  $\Delta\tau_n^2$ is
the time taken for $\psi_{x_0}$ to go backwards in time from
 $\Sigma_3$ to $\Sigma_0$ on the $n^{th}$ cycle we have
\begin{equation}\label{in:3+}
|\Delta\tau_n^2|<\tau_{2}
\end{equation}  
For some finite number $\tau_{2}$.
 
\smallskip

 Now, Since $\psi_{x_0}$ is incident on $\Sigma_0$ and hence,
 initially flows through $\Omega_\epsilon^+$, it follows from the inequalities
(\ref{in:1}), (\ref{in:2}), (\ref{in:3}) and (\ref{in:3+}) that
\[
\frac{\int_{I_c}d\tau}{\int_{I_b}d\tau}> \frac{1}{\tau_\epsilon (\alpha+1)}
                                        \ln\left(\frac{\epsilon}{p_0}\right)
\]
where $\tau_\epsilon=\max (\tau_{1}, \tau_{2}) $. Choose
\[
p_0= \epsilon\exp\left[-\frac{\tau_\epsilon (\alpha+1)}{
                             (\frac{m}{n}-1)}\right],
\]
 then we have
\begin{eqnarray}
\int_{I}d\tau&=&\int_{I_c}d\tau\left( 1 +\frac{\int_{I_m}d\tau}{\int_{I_c}d\tau}\right)\nonumber\\
 -\tau_f &<&\frac{m}{n}\int_{I_c}d\tau .\label{in:imp}
\end{eqnarray}
Combining (\ref{in:imp}), (\ref{int:pis}) and (\ref{eq:htime}) gives
\[
H(\tau_f)< H_0 e^{4n\tau_f}
\]
which is the required result.$\Box$
\\

We say a scalar field cosmology is non-trivial if there exists
some space-time point for which $\dot{\phi}$ is non-zero.  

\begin{theorem}
If $(g_{\mu\nu},\phi)$ is a non-trivial scalar field cosmology
with potential (\ref{eq:epot}) and if
$g_{\mu\nu}$ is spatially flat and isotropic, then $g_{\mu\nu}$ possesses
an initial space-time singularity.
\end{theorem}
Proof:

By Lemma \ref{lem:1} there exists $\epsilon$ and $p_m$ such that
 all trajectories of $(\ref{eq:S2exp})$ incident on $\Sigma_0$ with $p_0<p_m$ satisfy
\[
H(\tau)< H_0e^{2\tau}
\]
for all $\tau<0$. It will be sufficient to show that these trajectories 
reach the boundary in finite proper time. The proper time taken
to reach the boundary is given by
\begin{eqnarray}
\Delta t&=&\int_{-\infty}^0 x d\tau\\
        &<& x_0\int_{-\infty}^0 e^{\tau/2}d\tau\\
        &=&2x_0
\end{eqnarray}
Since this is finite, we have the result.$\Box$\\

\bigskip
\begin{theorem}

If $(g_{\mu\nu},\phi)$ is a non-trivial scalar field cosmology
with potential (\ref{eq:epot}) and if
$g_{\mu\nu}$ is spatially flat and isotropic, then particle horizons exist
for all isotropic observers (observers whose worldlines are tangent to the
timelike Killing vector field $U^\mu$).
\end{theorem}
Proof:
A particle horizon exists for an isotropic  observer at time $t$ if 
the integral
\[
l=\int_0^t \frac{1}{a} dt
\]
exists and is finite.
By the definition of $\tau$;
\[
a=e^{\frac{2}{3}\tau}
\]
and 
\[
\frac{dt}{d\tau}=x
\]
Therefore,
\[
l=\int_{-\infty}^{\tau} e^{-\frac{2}{3}\tau}x d\tau.
\]
By Lemma \ref{lem:1} there exists $\epsilon$ and $p_m$ such that
 all trajectories of $(\ref{eq:S2exp})$ incident on $\Sigma_0$ with
 $p_0<p_m$ satisfy
\begin{equation}\label{eq:horproof}
H(\tau)< H_0e^{\frac{5}{6}\tau}
\end{equation}
for all $\tau<0$. If $l$ is finite at $\tau=0$ it will be finite for all
$\tau$. It will therefore  be sufficient to show that trajectories 
for which (\ref{eq:horproof})  holds possess a horizon 
at $\tau= 0$. Using (\ref{eq:horproof}) we have 
\begin{eqnarray}
      l  &\leq& x_0\int_{-\infty}^0 e^{\frac{\tau}{6}}d\tau\\
        &=&6x_0
\end{eqnarray}
Since this is always finite we have the result.$\Box$ \\

\bigskip
Note also that as $x_0\rightarrow 0$, $l$ must also approach 0 indicating that
the horizon length shrinks to zero as $t\rightarrow 0$.

The physical meaning of the Lemma~\ref{lem:1} becomes clearer when we
recall that $\tau=\ln v$ and, by (\ref{eq:hx}) and the definition of $x$
$H=9K^{-4}$. Lemma~\ref{lem:1} may thus be interpreted as saying that 
for all $0<n<1$ there exists some $A>0$ such that
$$
K>Av^{-n}.
$$ 
Recalling that $K=\dot{v}/v$ it thus  follows that for all $m>1$ there exists $v_0$ such that 
\begin{equation}\label{vupper}
v(t) > v_0t^m
\end{equation}
Thus, in the neighbourhood of the singularity the volume element expands
faster than any power law with power greater than one. In order to avoid a
particle horizon it must expand slower than $t^3$.

What about the case $m=1$? Consider the function $e^{-4\tau}H$.
Taking the derivative of this function with respect to $\tau$, using 
(\ref{eq:hdiff})
$$
\frac{d\phantom{b}}{d\tau}e^{-4\tau}H = -4e^{-4\tau}H(1-q^2)
$$
Thus for $\tau<0$ we have 
$$
\ln\left(e^{-4\tau}H\right) = c + 4\int_\tau^0(1-q^2)d\tau
$$
where $c$ is a constant. The integral on the right hand side tends to infinity
as $\tau\rightarrow -\infty$ since each solution spends a finitely large
 amount of $\tau$-time with, say, $q^2<\epsilon$ on {\em each} cycle (of which there are an infinite number).

Thus,
$$
\lim_{\tau\rightarrow -\infty}e^{-4\tau}H =\infty. 
$$
This translates to a corresponding limit for $v$ and $t$, as above:
\begin{equation}
\lim_{t\rightarrow 0}t^{-1}v =0.
\end{equation}   
Comparing this expression with (\ref{vupper}) we see that $v$ expands
slower than $t$ but faster than any power law $t^p$ with $p>1$. 
In this respect the behavior of the gravitational field near the singularity is quite subtle and unusual since it can not be adequately 
modeled by a power law. Note also that it is clearly
 not admissible to neglect the dynamical effect of the potential when considering the gravitational field near the singularity.
 
\section{Conclusions}
The above example indicates that non-trivial asymptotic behavior can emerge, even in very simple models, as a result of  very steep self interaction
 potentials. It is of interest that both space-time singularity and particle horizons exist even
 though  the strong energy condition is violated by typical solutions 
during all periods of their evolution, including asymptotically close to the
singularity. 
     
It is suggested by the author that the qualitative features exhibited above,
including the oscillatory behavior and existence of a singularity and
and particle horizons,  are 
characteristic of FRW scalar field models which have very steep potential wells. Preliminary 
investigations of the steeper than exponential potential 
$V(\phi)=e^{\lambda\phi^2}$ and the ``hard wall'' potential 
$V(\phi)= 1/(\lambda -\phi^2)$ on the domain
 $\phi^2<\lambda $  reveal
 no significant departure
from the qualitative behavior of the exponential potential well.  
   
The asymptotic oscillatory behavior of the 
scalar field  is reminiscent of the oscillatory behavior of
the components of the shear tensor displayed by  the Belinskii, Khalatnikov, Lifschitz perfect fluid cosmologies \cite{bellip1,belip1.5,belip2}. This behavior is associated with the existence of dynamical
chaos in the solution space of these cosmologies \cite{barragain}.
 The apparent similarity suggests that it might be particularly  interesting to
investigate
the dynamics of exponential potential well models in slightly more complicated space-times such as Bianchi type IX. It seems not inconcievable that the coupling of scalar field and shear degrees of freedom could increase the degree of mixing, thereby resulting in partical horizon avoidance, particularly since, unlike the BKL solutions, the expansion of scalar field cosmologies is not constrained by the strong energy condition.

Another interesting feature of the solutions is that there exists two  natural time scales, namely the affine parameter
time $t$ and the period  of oscillation of the scalar field, which we might choose to characterise 
by some coordinate 
$\eta$. The timelike
geodesics are incomplete with respect to $t$ but complete with respect 
$\eta$.

 By convention we say that a space-time singularity exists if geodesics
are incomplete with respect to  their affine parameter
since this describes a situation where a freely falling observer would 
reach the edge of space-time in finite proper time. 
 However, the 
physical interpretation of $t$ as the proper time
 is not really meaningful on a neighbourhood of a boundary point 
 of space-time 
 since no normal coordinates can be constructed at such a point (normal coordinates can be constructed at a point arbitrarily close to a singularity but may never be extended to the singularity itself). 

On the other hand, $\eta$ can be interpreted as
 the most natural time scale associated with 
the matter content of the universe (including all clocks and astronauts 
since scalar matter is the only matter entering into this model) and therefore
might be a more meaningful measure of the time taken to reach the singularity.
One could imagine that if a more realistic cosmological model could be shown 
to
 display a similar oscillatory behavior
for its physical fields then one would be
 led to an interpretation whereby a 
singularity exists but can never be reached in a finite amount of time
by any physical object.

\begin{center}
{\large \bf Acknowledgments}
\end{center} 
The author would particularly like to thank Peter Szekeres for  helpful comments and suggestions on the content and layout of this article. This research was supported in part by an Australian Post Graduate Research
 Award.

\end{document}